\documentclass[12pt,fleqn]{article}

\usepackage{graphicx}
\usepackage{dcolumn}
\usepackage{bm}
\usepackage{amsmath}
\usepackage{amssymb} 
\usepackage{subfigure}
\usepackage{psfig,float}
\def\OMIT#1{}
\begin{document}

\title{A Constructive Approach to Gene Expression Dynamics}

\author{T. Ochiai{\footnote{These authors contributed equally to this work.}}, J.C. Nacher{\footnotemark[1]}, T. Akutsu}

\maketitle

\begin{center}
{\it Bioinformatics Center, Institute for Chemical Research, Kyoto University, }\end{center}
\begin{center}
{\it Uji, 611-0011, Japan}
\end{center}

\begin{center}
\end{center} 

\begin{center}
\end{center}

\begin{abstract}
{\small{The advent of new experimental genomic technologies and the massive increase of DNA sequence information
is helping researchers better understand how our genes work. Recently, experiments on mRNA abundance (gene expression) 
have revealed that gene expression shows a stationary organization described by a 
power-law distribution (scale-free organization) (i.e., gene expression $k$ decays as $k^{-\gamma}$), 
which is highly conserved in all the major five kingdoms of life, from Bacteria to Human. An underlying gene expression 
dynamics "rich-travel-more" was suggested to recover that evolutional conservation of transcriptional organization. Here we propose
a constructive approach to gene expression dynamics with larger scope. Our gene expression construction restores 
the stationary state, predicts 
the power-law exponent for different organisms with natural explanation for small correction at high and low expression levels, describes the intermediate state dynamics (time finite) and elucidates 
the gene expression stability. This approach requires only one assumption: Markov property.}}
\end{abstract}


\section{Introduction}

DNA encodes tens of thousands genes (around 30.000 genes in a human cell), which 
can be expressed as messenger ribonuclei acid (mRNA) transcripts and then translated into protein.
Proteins are essential for most biochemical processes in the cell and execute almost all cell functions. Therefore, the protein study
is one of the central issues in our post-genomic era.

Protein abundance in a cell depends on many factors. One of them is whether the respective gene is expressed (transcribed) or not, 
and how fast it is expressed. However, direct measures of protein abundance are difficult technically. With the advent of
new experimental genomic technologies as micrroarray chips, simultaneous expression levels of thousands of genes can be measured
and monitored under different conditions \cite{hart,david}. Though the relationship between gene expression and abundance of proteins in the cells
is not completely determined, it is possible to make precise estimations about the presence of proteins from gene expressions measures. Hence, 
the experimental study and theoretical analysis of gene expression organization in different organisms 
\cite{kaneko,kuznetsov}, and its underlying dynamics are important in biology. 

Evolutional gene expression organization has recently been studied by measuring the mRNA abundance (gene expression) for tens of thousands
of genes in parallel (GeneChips arrays) through several related experiments \cite{ueda}. From {\it E. Coli} to {\it Homo sapies}, the gene
distribution $p(k)$ (frequency of genes that have an amount of expression $k$) was revealed as a
stationary scale-free organization \cite{ueda} (i.e., $k$ decays as a power-law $k^{-\gamma}$ \cite{bara,amaral} ). This organization can be recovered by means of the 
underlying dynamics "rich-travel-more" explained in \cite{ueda} after some assumptions are considered. 

Here, in this paper, we propose an
innovative construction of gene expression dynamics, which {\it spontaneously} re-builds the observed stationary power-law organization, by making 
use of only one natural assumption: gene expression has "short memory" (Markov property). In addition, our gene expression 
construction predicts the power-law exponent for different organisms with natural explanation for small correction at high and low expression levels, describes the intermediate state dynamics (time finite) and elucidates 
the gene expression stability.

The paper is organised as follows. In Section 2, we describe the methods (subsection 2.1) 
and present our results (subsection 2.2). In Section 3, we summarize our work.

\section{ Methods and Result}
\subsection{Methods}
\subsubsection{Natural construction}
\paragraph{Why Markov property.}
If we say that the system has a Markov property, we mean that the future is governed by the present and does not depend on the past. Our model is based on Markov property. This 
assumption is very natural for biology since 
all biological systems obey the physical laws, which manifest Markov property.  
 
\paragraph{Why probability.} 
On the other hand, it seems very natural to use stochastic models because 
cells are so complex systems that we can not precisely analyse all the variables. Therefore, it 
is not sufficient to use only deterministic models and we must approach to the problem by using stochastic processes theory.

\begin{figure}[htb]
\setlength{\unitlength}{1cm}
\begin{picture}(15,10)(-1,-1)
\put(-2,0
){\includegraphics{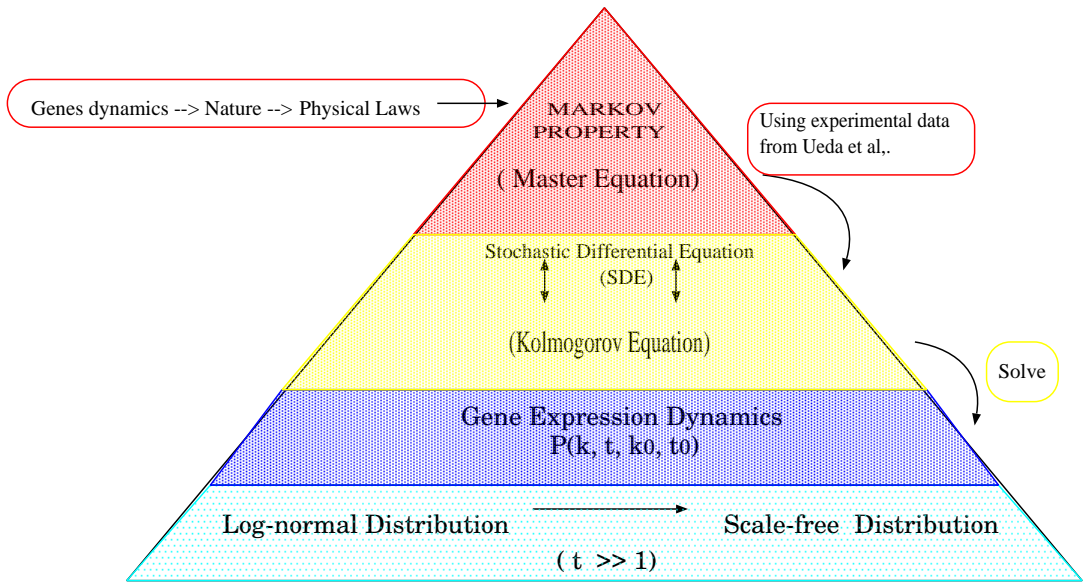}}
\end{picture}
\caption{\small{Scheme of the fundamental levels of our construction. From 
Markov property (or equivalently Master Equation)
to scale-free distribution.}}
\label{fig: construction}
\end{figure}

In Fig. \ref{fig: construction} we sketch our construction at different level of knowledge. Next subsections explain in detail each pyramidal level, from general 
and fundamental concepts to particular and operative solutions (stationary organization and dynamics).
\subsubsection{Markov property and Master equation.}

\paragraph{Markov property.}
Let $\{X_t, 0 \le t <\infty \}$ be a stochastic process. For $(t_n>\cdots>t_0)$,  the conditional probability density function
\begin{eqnarray*} 
p(x_n,t_n|x_{n-1},t_{n-1};\cdots;x_0,t_0)=p(X_{t_n}=x_n | X_{t_{n-1}}=x_{n-1} ;\cdots; X_{t_0}=x_0 )
\end{eqnarray*}
is defined as usual manner. It is said that a stochastic process has "Markov property", when the condition
\begin{eqnarray}\label{eqn: Markov property}
p(x_n,t_n|x_{n-1},t_{n-1};\cdots;x_0,t_0)=p(x_n,t_n|x_{n-1},t_{n-1}) 
\end{eqnarray}
holds for arbitrary $t_n>\cdots>t_0$. Roughly speaking, Markov process means that the future does not depend on the past, but only on the present time. In what follows, we assume that the probability density $p(x,t|x_0,t_0)$ has the time  translation invariance $p(x,t|x_0,t_0)=p(x,t+a|x_0,t_0+a)$ for arbitrary $a$.

Our only one assumption is that "gene expression has the Markov property". More precisely, we assume 
that the gene expression level 
is denoted by the stochastic process with Markov property $X_t$. This assumption is quite natural, because the gene 
expression system belongs to nature, the nature obeys the physical laws 
and the physical laws manifest the Markov property (see Fig. 1).

\paragraph{Master equation.}
For the matter of convenience, we write $p(x,t)$ for $p(x,t|x_0,t_0)$. Then, if the stochastic 
process has the Markov property (Eq. (\ref{eqn: Markov property})), the conditional probability density 
function $p(x,t|x_0,t_0)$ obeys Kramers-Moyal expansion (which is mathematically equivalent to the Master equation):
\begin{eqnarray}\label{eqn: master equation}
\frac{\partial p(x,t)}{\partial t}=\sum_{n=1}^{\infty}\frac{(-1)^n}{n!}\frac{\partial^n}{\partial x^n}\{a_n(x)p(x,t)\},
\end{eqnarray}
where
\begin{eqnarray}\label{eqn: initial condition}
a_n(x)=\lim_{\epsilon \to 0}\frac{1}{\epsilon}\int (y-x)^n T_\epsilon(y,x) dy.
\end{eqnarray}
Here $T_\epsilon(y,x)$ is an instantaneous transition probability defined by $T_\epsilon(y,x)=p(y,t + \epsilon | x,t)$ for sufficiently small $\epsilon$. Details of the proof can be found in Ref. \cite{Kampen}. 

In the context of gene expression level, this equation (\ref{eqn: master equation}) represents the dynamics of abundance of mRNA (gene expression). 

\paragraph{How to use Master equation.}

We can obtain the dynamics of probability density $p(x, t ~|~ x_0, t_0)$ for any time $t$, from 
experimental data of instantaneous transition probability $T_\epsilon(y,x)$ ($\epsilon$ is sufficiently small and fixed), by the following procedure:

\begin{itemize}

\item[{\it (i)}] Given the experimental data of instantaneous transition probability $T_\epsilon(y,x)$ ($\epsilon$ is sufficiently small), we obtain $a_n(x)$ by using Eq. (\ref{eqn: initial condition}).\footnote{Notice that we do not need the whole data $T_\epsilon(y,x)$ for any $\epsilon$. The necessary data is $T_\epsilon(y,x)$ at a sufficiently small fixed $\epsilon$.} 

\item[{\it (ii)}] By inserting $a_n(x)$ into Kramers-Moyal expansion (\ref{eqn: master equation}) (Master equation) and by solving this PDE 
(partial differential equation), we can obtain $p(x,t~|~x_0,t_0)$. 
\end{itemize}

In general, it is difficult to solve Kramers-Moyal expansion (Eq.(\ref{eqn: master equation})) (Master equation) 
in the above second process {\it (ii)}. However, for particular cases we can solve this equation analytically, which will be explained in next sections.

In the sequel, we use notations $k_i$ (i=1,2) for gene expression levels instead of $x, y$. For example, $T_\epsilon(k_2,k_1)$ 
denotes the instantaneous transition probability of expression level from $k_1$ to $k_2$.

\subsubsection{Initial instantaneous transition data $T_\epsilon(k_2,k_1)$}
\label{sec:initial condition } 
\begin{figure}[htb]
\setlength{\unitlength}{1cm}
\begin{picture}(15,12)(-1,-1)
\put(-4,0){\includegraphics{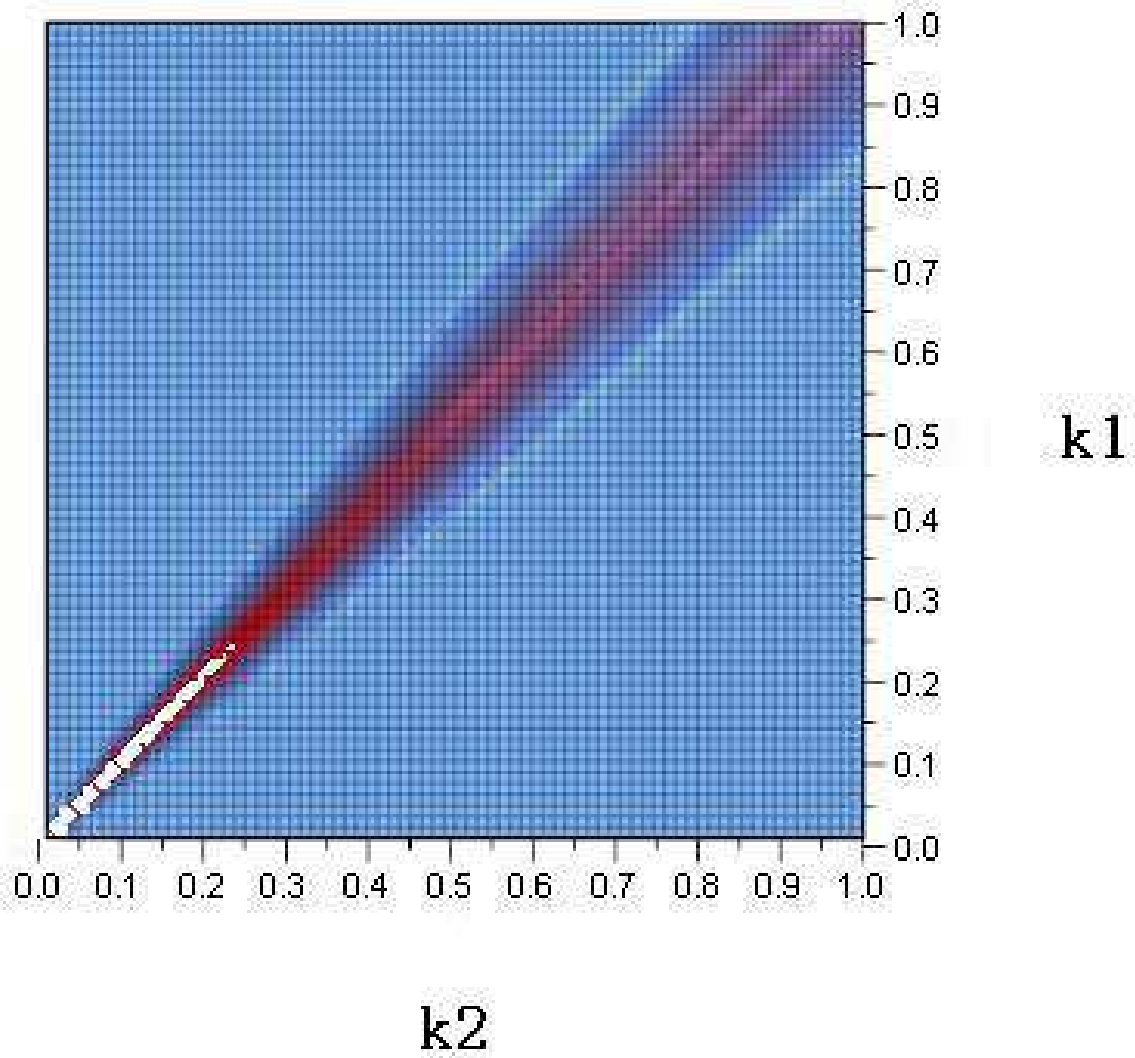}}
\end{picture}
\caption{\small{Transition probability between two states of gene expression $T_{\epsilon}(k_2, k_1)$. We use analytical expression for reproducing
qualitatively the same behaviour observed experimentally by \cite{ueda}. Colors represent gradual changes of 
values of transition probability: white (maximum), red (medium), light blue (minimum). }}
\label{fig:Initial daata}
\end{figure}

Recently, gene expression instantaneous transition probability (ITP) $T_\epsilon(k_2,k_1)$ for individual 
genes, from expression level $k_1$ to expression level 
$k_2$ along different conditions, was obtained experimentally \cite{ueda}.
The following expression
\begin{eqnarray}\label{eqn:initial data T} 
T_\epsilon(k_2,k_1)=\frac{1}{\sqrt{2\pi(\sigma k_2)^2\epsilon}} \exp\Bigl[-\frac{(\log(k_2/k_1)-(\mu-\frac{1}{2}\sigma^2)\epsilon)^2}{2\sigma^2\epsilon}\Bigr],  
\end{eqnarray}
with $\mu=-0.07$, $\sigma=0.25$ and $\epsilon=0.01$, faithfully reproduces 
the experimental data of ITP of {\it S.Cerevisiae}, shown in \cite{ueda}.  
Fig. \ref{fig:Initial daata} shows this ITP $T_\epsilon(k_2,k_1)$. It 
represents how expression level $k_1$ (horizontal axis) changes into expression level $k_2$ (vertical axis)
 under a small time interval. Colors represent gradual changes of values of instantaneous transition probability, from 
 red (maximum) to light blue (minimum).

In what follows, our aim is to compute probability density $p(k,t)$ from the input data $T_{\epsilon}(k_2,k_1)$, using the procedure given in the previous section. 

\subsubsection{Computation of $a_n(k_1)$}
In this section, we compute $a_n(k_1)$ from the initial data of ITP $T_\epsilon(k_2,k_1)$. We insert Eq. (\ref{eqn:initial data T}) into 
\begin{eqnarray}
a_n^{\epsilon} (k_1)=\frac{1}{\epsilon}\int (k_2-k_1)^n T_\epsilon(k_2,k_1) dk_2.
\end{eqnarray}
Then, after some computation, we have
\begin{eqnarray}
a_n^{\epsilon} (k_1)=(k_1)^n \int^\infty_{-\infty}
\frac{dz}{\sqrt{\pi}}
e^{-z^2}\frac{1}{\epsilon}(z\sqrt{2\sigma^2}\sqrt{\epsilon}+(\mu-\frac{\sigma^2}{2}\sigma^2z^2)\epsilon+O(\epsilon^{3/2}))^n.
\end{eqnarray}
After we take limit $\epsilon \to 0$, we finally obtain:
\begin{eqnarray*}
&&a_1(k_1)=\mu k_1, \\
&&a_2(k_1)=(\sigma k_1)^2,\\
&&a_i(k_1)=0 ~~~~~~ (i\ge 3).
\end{eqnarray*}
Here we remark the following. Although we have fixed the value of $\epsilon$ at $\epsilon=0.01$ in Eq. (\ref{eqn:initial data T}), the above limiting procedure $\epsilon \to 0$ is still valid as a sufficiently good approximation. Futhermore, it is remarkable that all $a_i$ $(i=3,4,\cdots)$ vanish.

\subsubsection{Emergence of Kolmogorov equation and SPDE}
In the last section, we find out that $a_1(k_1)=\mu k_1$, $a_2(k_1)=(\sigma k_1)^2$ and $a_i(k_1)=0$ $(i\ge 3)$, where $\mu=-0.07$ and $\sigma=0.25$. However, in order to keep the argument more general, we still consider $a_1(k_1)$ and $a_2(k_1)$ arbitrary while we assume that $a_i=0$ vanish for $i\ge 3$.

\paragraph{Kolmogorov Equation.}
If $a_i(k_1)=0$ vanish for $i\ge 3$, then Kramers-Moyal expansion (\ref{eqn: master equation}) (Master equation) becomes Kolmogorov equation:       

\begin{eqnarray}
\frac{\partial p(k,t)}{\partial t}=-\frac{\partial}{\partial x}\{a_1(k)p(k,t)\}
+\frac{1}{2}\frac{\partial^2}{\partial k^2}\{a_2(k)p(k,t)\},
\end{eqnarray}
where $p(k,t)=p(k,t|k_0,t_0)$.

\paragraph{SPDE.}
In addition, it is known that the Kolmogorov equation 
is equivalent to the following stochastic partial differential equation (SPDE):

\begin{eqnarray}\label{eqn: SPDE}
dX_t=\alpha(X_t) dt + \beta(X_t)dW_t,
\end{eqnarray}
where stochastic variable $X_t$ denotes the gene expression level, $\alpha(X_t)=a_1(X_t)$ denotes the instantaneous 
transition of the average of the gene expression level per unit time, $\beta(X_t)=\sqrt{a_2(X_t)}$ denotes the 
instantaneous transition of variance of the gene expression level per unit time  and $W_t$ denotes the Wiener process \cite{Wong}. 

\subsubsection{Analysis of our model}
From general analysis of Kolmogorov equation, we return to our original situation where $a_1(k_1)=\mu k_1$, $a_2(k_1)=(\sigma k_1)^2$ and
$a_i(k_1)=0$ vanish for $i\ge 3$. Then, by imposing the condition $\alpha(X_t)=\mu X_t$ and $\beta(X_t)=\sigma X_t$ 
on the SPDE (Eq. (\ref{eqn: SPDE})), we obtain   

\begin{eqnarray}\label{eqn: BS}
dX_t=\mu X_t dt + \sigma X_tdW_t.
\end{eqnarray}
This reduced model is also known as Black-Scholes model in financial engineering (see Ref. \cite{black}). This SPDE 
directly gives useful information about the properties of the model as follows:

\begin{itemize}

\item[{\it (i)}] {\it Negative $\mu$}. From subsection \ref{sec:initial condition }, we know that $\mu$ is negative. This means 
that the gene expression level has a slightly decreasing tendency, which seems natural for biological systems. 


\item[(2)] {\it Rich-travel-more}. This mechanism, which regenerates the stationary power-laws, 
was introduced by \cite{ueda}. In our construction, this mechanism is not concealed at all, and it is evident just by 
looking at the second term $(\sigma X_tdW_t)$ in Eq. (\ref{eqn: BS}).   
\end{itemize}

\subsubsection{Gene expression dynamical solution}

By using Ito formula, we can solve the SPDE (Eq. (\ref{eqn: BS})) and derive the dynamics of gene expression (the {\it expression-temporal} probability density )

\begin{eqnarray}\label{eqn: final solution}
p(k,t|k_0,t_0)=\frac{1}{\sqrt{2\pi(\sigma k)^2(t-t_0)}}\exp\Bigl[-\frac{(\log(k/k_0)-(\mu-\frac{1}{2}\sigma^2)(t-t_0))^2}{2\sigma^2(t-t_0)}\Bigr],
\end{eqnarray}
where $\mu=-0.07$ and $\sigma=0.25$.\footnote{Certainly, we can formally 
obtain this solution (Eq. (\ref{eqn: final solution})) by substituting the initial 
instantaneous transition 
probability $T_\epsilon(k_1,k_2)$ (Eq. (\ref{eqn:initial data T}) ) into $p(y,t + \epsilon | x,t)=T_\epsilon(y,x)$. However, this 
argument is not correct.}
From this 
equation, we can compute the probability of finding that the gene expression level takes the value  $k$ at arbitrary time $t$. This equation explains how the gene expression behaves as a function of time. This is our main result.

\subsection{Interpretation of results}

\subsubsection{Results when time is infinity}

The following results are obtained as asymptotic behavior of the system for large time:

{\it (i) Model shows an almost stationary state :} In Eq. (\ref{eqn: final solution}), if we 
take limit $t\to \infty$, then we re-build the power-law distribution observed experimentally
\begin{equation}\label{eqn: sacle free}
p(k)\propto k^{\frac{2\mu/\sigma^2-5}{4}},
\end{equation}
where $p(k)=\lim_{t\to \infty} p(k,t|k_0,t_0)$. Here, we remark that $p(k,t|k_0,t_0)$ essentially does not depend on both $k_0$ and $t_0$ if we take limit $t\to \infty$. The result indicates how the system is organized 
in a stationary state after long time\footnote{This result Eq. (\ref{eqn: sacle free}) is different
 from that of \cite{ueda}. It is because they assume $\mu=0$ and $\frac{\partial p(x,t)}{\partial t} = 0$ in \cite{ueda}. In our approach, $\mu$ is obtained from experimental data and we do not assume $\frac{\partial p(x,t)}{\partial t} = 0$}. 
 
\begin{figure}[htb]
\setlength{\unitlength}{1cm}
\begin{picture}(15,10)(-1,-1)
\put(-4,0){\includegraphics{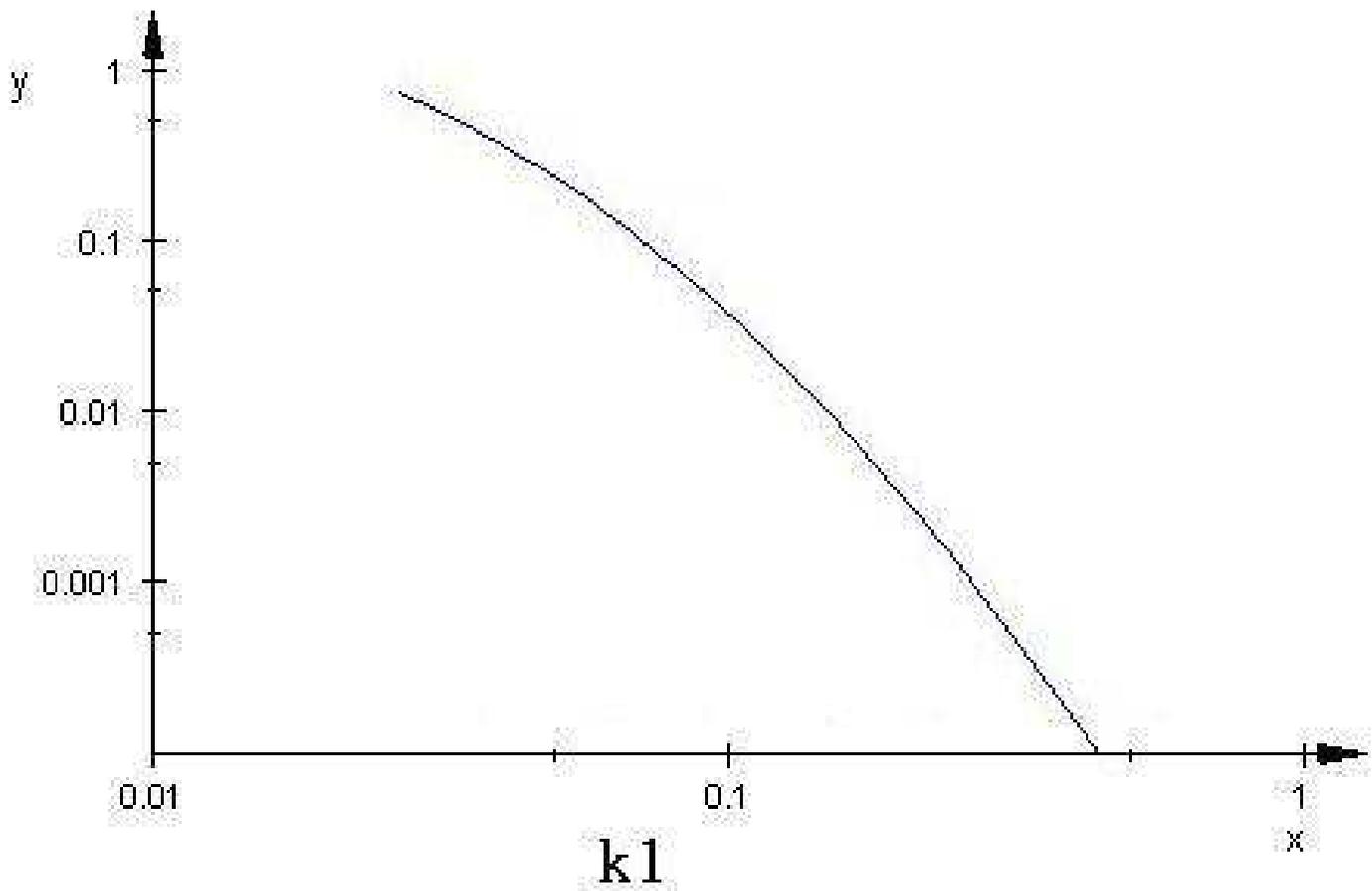}}
\end{picture}
\caption{\small{Gene distribution $p(k)$ in vertical axis in terms of gene expression level $k$ in horizontal axis. Scale is log-log. We 
plot the stationary state corresponding to the Log-Normal distribution. As we see the distribution is almost scale-free, with small
curvature correction. }}
\label{fig:power law}
\end{figure}

 Once that scale-free organization has been reached, it becomes stationary. However, it 
is worth noticing that when the system continues evolving (time passing), the absolute values of gene expression level still continue 
decreasing or increasing with the time in order to keep the total probability equal to one. However, the global scale-free organization is invariant. Due to this reason, we also call  
it {\it almost} stationary state. We can see the probability distribution in Fig. \ref{fig:power law}.

{\it (ii) Model predicts power-law organization :} It includes prediction of the scale-free exponent $\gamma$ for each
different organism (or dataset). As we see in Eq. (\ref{eqn: sacle free}), the expression of $\gamma$ is $-{\frac{2\mu/\sigma^2-5}{4}}$. By knowing
$\mu$ and $\sigma$ from experimental data, it is straightforward to obtain the respective value of $\gamma$. In particular, we have obtained
$\gamma=1.81$ for the organism {\it S. Cerevisae} in good agreement with experimental results in \cite{ueda}, by using $\mu=-0.07$ and
$\sigma=0.25$. The same procedure can be applied for obtaining $\gamma$ values for different organisms.

{\it (iii) Model predicts small corrections in the scale-free distribution :} As we can see from Fig. \ref{fig:power law}, the results predicted by our
construction differ slightly from a scale-free distribution. It contains a small curvature, reflecting the original 
Log-Normal distribution. It is worth noticing that this curvature is also found at the experimental data from \cite{ueda} 
for low expression level (high probability) and large expression level (very low probability). Hence, the agreement
 with data from \cite{ueda} is very good.

{\it (iv) Model elucidates robustness (stability) :} It means that, by following our construction, any gene expression level $k_0$
at the initial time $t_0$ will evolve to the same stationary final state, self-organizing like a scale-free distribution. Hence, the final 
state is independent and stable under changes done at the initial state.

It is worth remarking that all these findings are supported by experimental data \cite{ueda}.

\subsubsection{Intermediate gene expression states (time finite)}

\begin{figure}[htb]
\setlength{\unitlength}{1cm}
\begin{picture}(15,15)(-1,-1)
\put(-5,0){\includegraphics{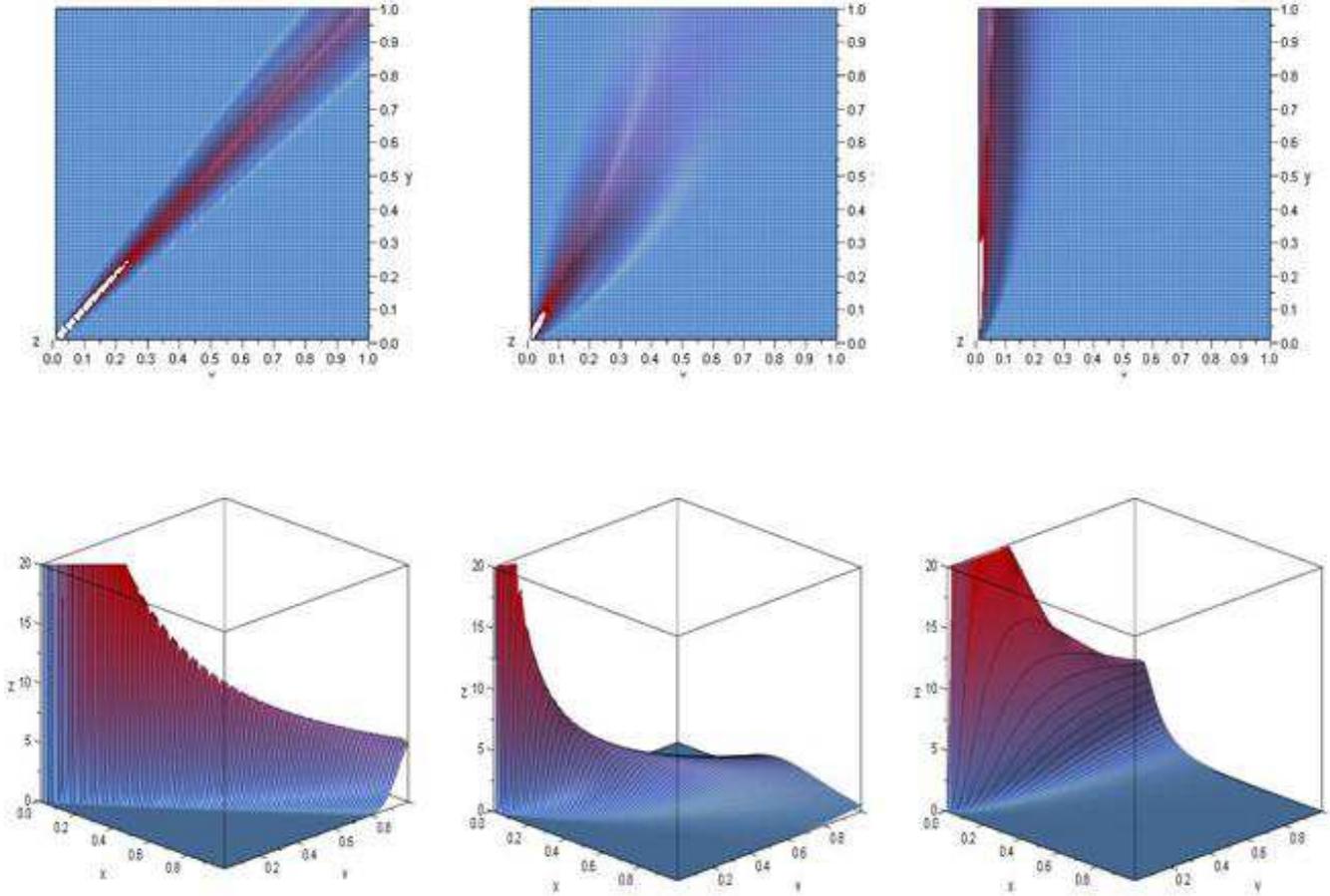}}
\end{picture}
\caption{\small{We show the dynamics of gene expression $p(k,t,k_0,t_0)$ 
(vertical axis). $y$ (resp. $x$) axis represents $k_1$ (resp. $k_2$) expression level. At the top, from left to right, three different
states: initial state ($t=0$), intermediate state (intermediate $t$), and final state ($t \to \infty$). At the bottom, the 
same representation by using 3D view. Colors represent gradual changes of 
values of transition probability: white (maximum), red (medium), light blue (minimum).}}
\label{fig:dynamics}
\end{figure}

From Eq. (\ref{eqn: final solution}), we can compute the probability of finding expression level $k$ at arbitrary time $t$. Namely, this equation explains how the gene expression changes over time. 
We show the evolving dynamics of gene expression in Fig. \ref{fig:dynamics} {\it (a, b, c)} (from left to right): {\it (a)} initial time, {\it (b)} intermediate time or finite fixed time (now the system is evolving), {\it (c)} 
 time is very large (final state). These results are in agreement 
 with the numerical simulated data shown in {\it Supplementary Material} 
 of ref. \cite{ueda}.

\section{Conclusions}

We have developed a constructive approach to gene expression dynamics based on only one theoretical assumption: Markow property.
Our gene expression construction restores the stationary state, predicts 
the power-law exponent for different organisms with natural explanation for small correction at high and low expression levels, describes the intermediate state dynamics (time finite) and elucidates 
the gene expression stability. Furthermore, our model is in agreement with the experimental data shown in \cite{ueda}.

Many extensive analyses have recently been done for studying the organization 
and topology of biological networks \cite{jeong, wagner, rasvaz}. However, a 
theoretical and rigorous analysis for explaining the dynamical complexity of these biological networks has not been developed.
An adequate and self-consistent theoretical framework would be illuminating and fruitful for understanding the cell dynamics.
 
Our construction may play the same role for the gene expression dynamics as the Schr\"{o}dinger equation in 
Quantum Mechanics or the Conservation of Energy and Newton's Law in Classical Mechanics. By following 
our construction, it seems plausible
to predict the future behaviour of a gene expression dynamical system. This marvelous gene expressions mechanics
may provide a better understanding of cell dynamics. As a future work, this approach may be extended for studying 
multi-gene correlations dynamics,
which hold interesting information of gene functionality.

\newpage

\newpage

\newpage@@

\newpage

\end{document}